\documentclass[fleqn,usenatbib]{mnras}

\usepackage{newtxtext,newtxmath}
\usepackage[T1]{fontenc}

\usepackage{graphicx}	% Including figure files
\usepackage{amsmath}	% Advanced maths commands
\title[Symmetry in Crab Pulsar's FLC]{Reflection Symmetry in the Folded Light Curve of the Crab Pulsar from NICER}

\author[M. Vivekanand]{
M. Vivekanand\thanks{E-mail: viv.maddali@gmail.com}
\\
No. $24$, NTI Layout $1$\textsuperscript{st} Stage, $3$\textsuperscript{rd} Main,
$1$\textsuperscript{st} Cross, Nagasettyhalli, Bangalore $560094$, India.
}

\date{Accepted 2022 May 5. Received 2022 April 16; in original form 2022 January 13}

\pubyear{2022}

\begin{document}
\label{firstpage}
\pagerange{\pageref{firstpage}--\pageref{lastpage}}
\maketitle

\begin{abstract}
	The Rotation powered pulsars Crab, Vela and Geminga have double peaked folded light curves 
	(FLC) at $\gamma$-ray energies, that have an approximate reflection symmetry. Here this 
	aspect is studied at soft X-ray energy by analyzing a high resolution FLC of the Crab pulsar 
	obtained at $1 - 10$ keV using the {\it{NICER}} observatory. The rising edge of the first 
	peak of the FLC and the reflected version of the falling edge of the second peak are 
	compared in several ways, and phase ranges are identified where the two curves are 
	statistically similar. The best matching occurs when the two peaks are aligned, but only in 
	a small phase range of $\approx 0.0244$ just below their peaks; their mean difference is 
	$-0.78 \pm 1.8$ photons/sec with a reduced $\chi^2$ of $0.93$. If the first curve is 
	convolved by a Laplace function, the corresponding numbers are phase range of $\approx 
	0.0274$, mean difference of $-1.23 \pm 1.30$ and $\chi^2$ of $0.76$. These phase ranges are 
	much smaller than those over which the reflection symmetry has been perceived. 
	Therefore the only way the two edges can have a mirror relation over a substantial phase 
	range is if one invokes a broad and faint emission component of amplitude $\approx 100$ 
	photons/sec and width $\approx 0.1$ in phase, centered at phase $\approx 0.1$ beyond the 
	second peak.
\end{abstract}

%
%________________________________________________________________

\section{Introduction}

\begin{figure}
\centering
\advance\leftskip-0.3cm
\includegraphics[width=9.2cm]{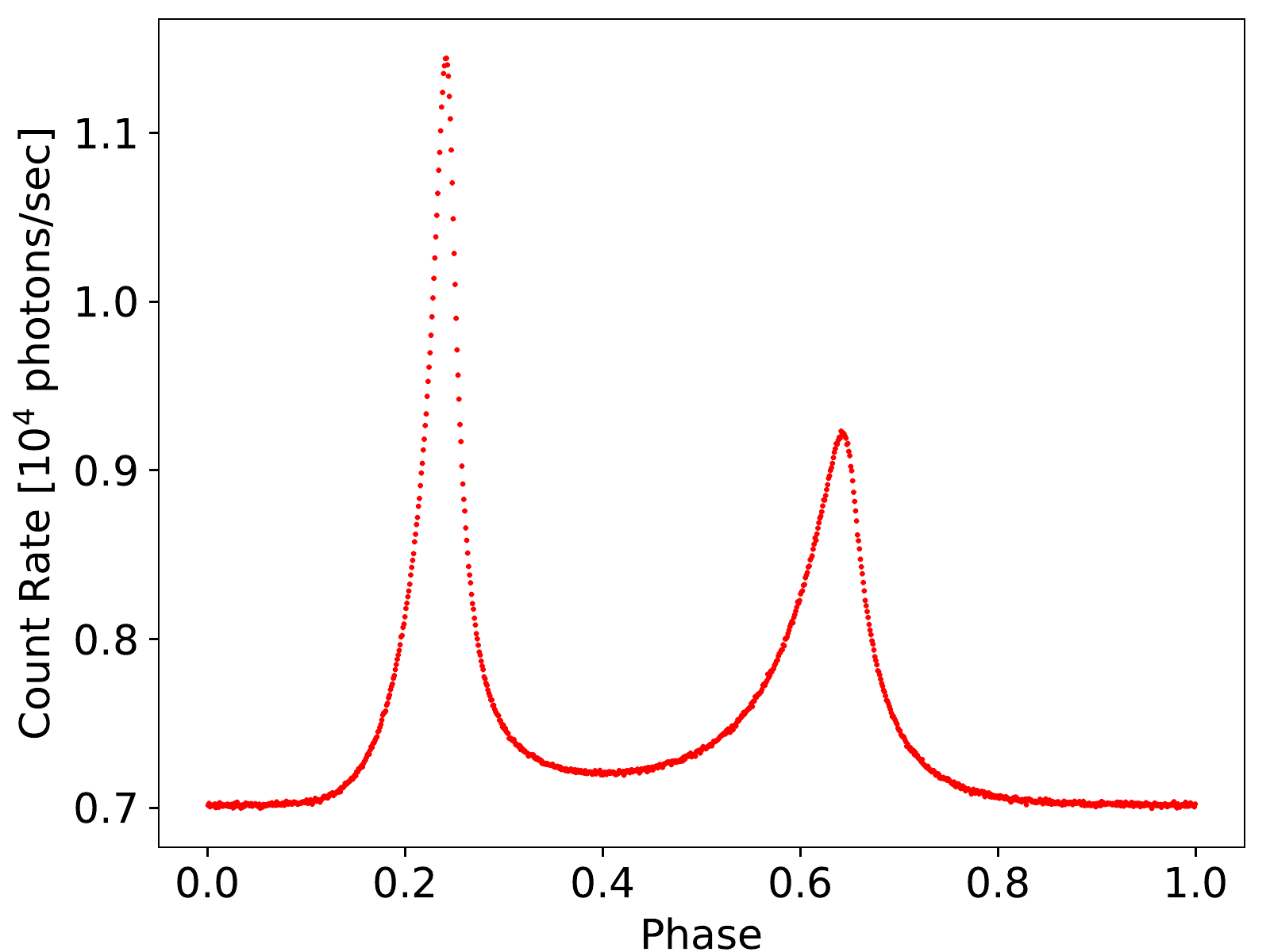}
\vskip-0.2cm
\caption{
	Crab pulsar's FLC in the energy range $1 - 10$ keV from {{\it NICER}}
	having $1024$ phase bins.
        }
\label{fig1}
\end{figure}

Figure~\ref{fig1} shows the folded light curve (FLC) of the Crab Pulsar in the soft X-ray 
energy range $1 - 10$ keV, obtained by \citet{Vivekanand2021} using the Neutron star 
Interior Composition Explorer ({\it{NICER}}) satellite observatory \citep{Arzoumanian2014, 
Gendreau2016}. It is similar to the FLC in Fig.~$15$ of \citet{Vivekanand2021} except that 
phase $0$ now represents the statistical minimum of the FLC, obtained by fitting a cubic 
polynomial to the first $275$ data of the earlier FLC. Visually it is conceivable that 
there is an approximate reflection symmetry between the two peaks in Fig.~\ref{fig1} 
provided their amplitudes are similar. Such a possibility has been evident to $\gamma$-ray 
astronomers in three of the brightest rotation powered pulsars (RPP) --- Vela \citep{Abdo2009}, 
Crab \citep{Abdo2010a} and Geminga \citep{Abdo2010b}, with Vela being the clearest example.

In the Vela pulsar's $\gamma$-ray FLC \citet{Abdo2009} note that its two peaks are 
asymmetric, and that the second peak has a slow rise and a fast fall, and the first 
peak has the opposite behavior (see their Sect. $4.1$, page $1088$). \citet{Bai2010} 
mention that Vela's FLC has a horn structure that is symmetric upon reflection around 
the middle of the bridge (see their Sect. $5$, page $1292$). \citet{Abdo2010a} note 
in their Sect. $4.1$ on page $1257$ that the two peaks of the Crab pulsar's $\gamma$-ray 
FLC are asymmetric, and that the second peak has a slow rise and a steeper fall. 
\citet{Saito2010} notes that the rising and falling edges of the two peaks in the 
$\gamma$-ray FLC of the Crab behave exponentially, and that the slopes are not 
symmetric between the rising and falling edges (Sect. $7.3$, page $70$). Although 
\citet{Abdo2010b} do not specifically mention any asymmetry in the two peaks 
of the $\gamma$-ray FLC of the Geminga pulsar, the argument of \citet{Bai2010} appears 
to hold for this RPP also (Fig.~$2$ of \citet{Abdo2010b}). \citet{Cheng2000} note that 
in earlier $\gamma$-ray observations of these three pulsars from {{\it EGRET}}, the two 
peaks are not symmetric. In their Sect. $4.4$ on page $972$, \citet{Cheng2000} 
specifically mention that their simulations of the Crab pulsar's FLC (in their Fig.~$7$) 
have some sort of symmetry. Thus several astronomers have noted the apparent asymmetry 
between the two peaks in the $\gamma$-ray FLCs of these three RPPs, giving the impression 
of a reflection symmetry between the two peaks.

This work explores this aspect in Fig.~\ref{fig1} above, the focus being on the reflection 
symmetry between the rising edge of the first peak and the falling edge of the second peak. 
Comparison of the falling edge of the first peak and the rising edge of the second peak 
would be model dependent due to the bridge emission, and therefore has not been pursued
here; this is elaborated upon later on.

The reflection symmetry noticed by earlier astronomers is essentially a visual effect. It 
was implied (in my opinion) over a phase range of the order of $\approx 0.2$ which would be 
approximately from phase $0$ to the first peak in Fig.~\ref{fig1}. To the best of my 
knowledge no one has so far studied these issues quantitatively. This work attempts a
quantitatively analysis of this subject and concludes that the reflection symmetry exists,
if at all, only in a small phase range close to the two peaks. However one can believe it
to exist over a much larger phase range if one is willing to accept the existence of
a weak and broad emission component in the wings of the second peak.

An important requirement for this work is the highest possible phase resolution in
Fig.~\ref{fig1} which has $1024$ phase bins per period, which is the highest 
currently available. This is possible because of the excellent photon collecting and 
timing ability of {{\it NICER}}. \citet{Vivekanand2021}
discusses the required accuracy on the rotation frequency of the Crab pulsar $\nu(t)$ 
and its time derivative $\dot \nu(t)$ as a function of epoch $t$, to obtain this 
phase resolution. Often $\gamma$-ray and X-ray FLCs of RPPs are obtained by using 
contemporaneous radio timing ephemeris. This should be good enough for most milli 
second pulsars whose $\nu(t)$ changes relatively slowly with epoch and which have 
low timing activity. However for the Crab pulsar the systematic differences between 
the $\nu(t)$ and $\dot \nu(t)$ estimated at radio and X-rays should be taken into 
account while forming high phase resolution FLC. The $\nu(t)$ and $\dot \nu(t)$ 
used to form Fig.~\ref{fig1} have been obtained self consistently from {{\it NICER}} 
data itself \citep{Vivekanand2020}, to yield an effective resolution of better than 
$1/512$ in phase \citep{Vivekanand2021}.

\section{Observations and data analysis}

The {{\it NICER}} observations of this work as well as their analysis  have been 
described in detail by \citet{Vivekanand2020, Vivekanand2021} including data 
calibration and filtering.

\begin{figure}
\centering
\advance\leftskip-0.3cm
\includegraphics[width=9.2cm]{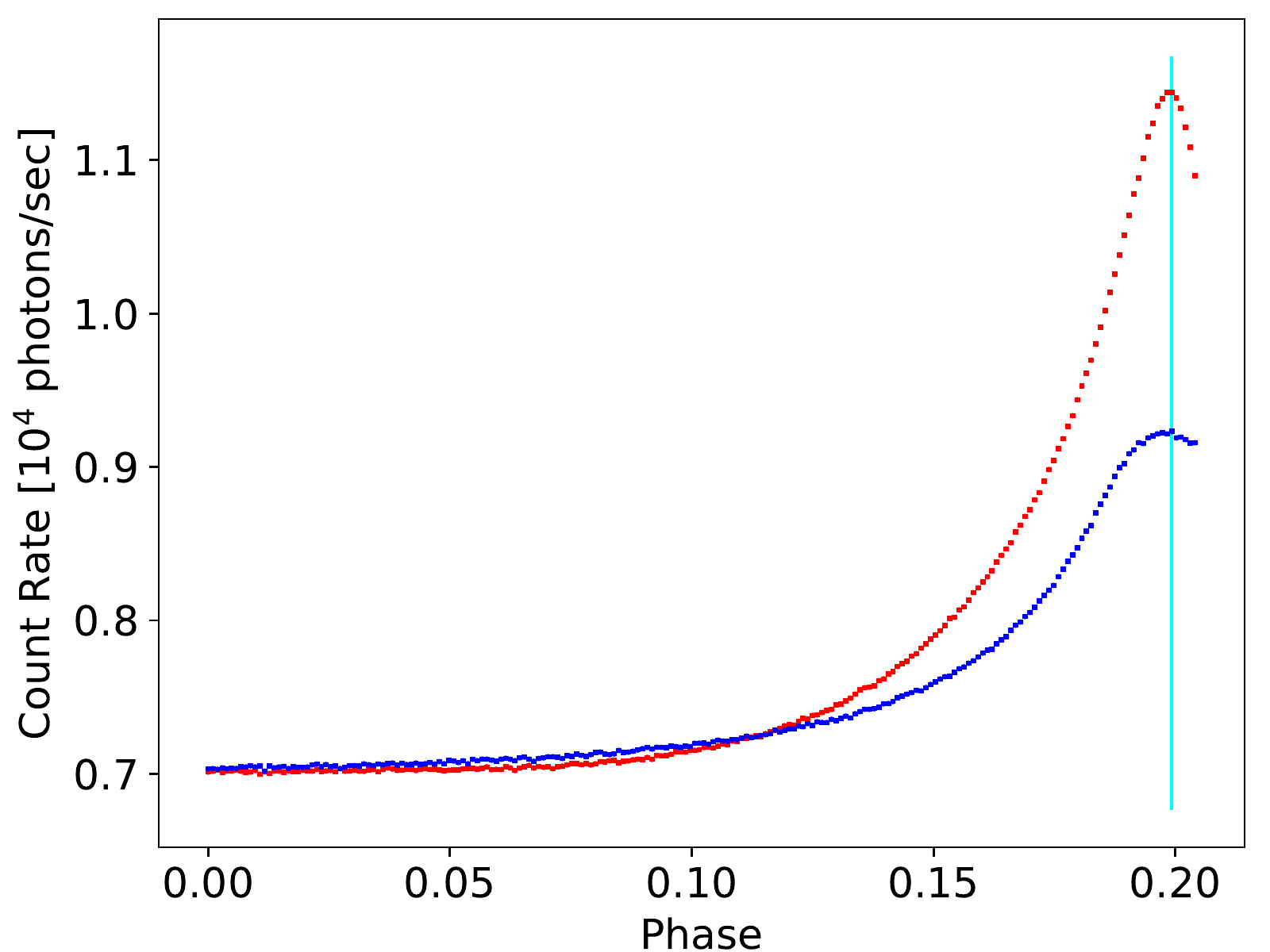}
\vskip-0.2cm
\caption{
	Superposition of the rising edge of the first peak (red) and reflected version
	of the falling edge of the second peak (blue), aligned at their peaks (cyan 
	line which is at phase $0.19921875$).
        }
\label{fig2}
\end{figure}

Fig.~\ref{fig2} shows the rising edge of the first peak of Fig.~\ref{fig1} and the 
reflected version of the falling edge of the second peak, aligned such that their peaks 
lie at the same phase. The red curve in Fig.~\ref{fig2} is identical to samples $1$ 
to $253$ of the FLC in Fig.~\ref{fig1}, while the blue curve is identical to samples 
$652$ to $1024$ after abscissa inversion; the peaks of the two curves at samples
$248$ and $657$ respectively are aligned at phase $0.19921875$  in Fig.~\ref{fig2} 
(cyan line), and only a smaller portion of the
curves has been plotted in the figure. The main effort in this work is 
to equalize the areas under the two curves so that their peak amplitudes are similar, 
and identify phase ranges where they match statistically. One searches for a good 
match in the space of four parameters: finer phase adjustment to the phase alignment 
of the two curves $\phi_0$; the initial and final phases of the phase range for 
matching $\phi_1$ and $\phi_2$; and a multiplicative correction to the relative area 
normalization of the two curves $\kappa$.  This will be known as analysis without 
any smoothing of the red curve.

However, Fig.~\ref{fig1} indicates that the second peak may be wider than the first peak. So
the above analysis is repeated after convolving the red curve with a smoothing function.
If the smoothing function is symmetric, like the Gaussian, an additional parameter is added
to the search space, which is a measure of the width $\sigma$ of the smoothing function.
This will be known as analysis with symmetric smoothing. If the smoothing function is 
asymmetric then two additional parameters are added to the search space, which are a measure 
of the widths $\sigma_{-}$ and $\sigma_{+}$ of the smoothing function on the negative and 
positive phase sides of its peak. This will be known as analysis with asymmetric smoothing. 

\section{Results of analysis without smoothing}

\begin{figure}
\centering
\advance\leftskip-0.3cm
\includegraphics[width=9.2cm]{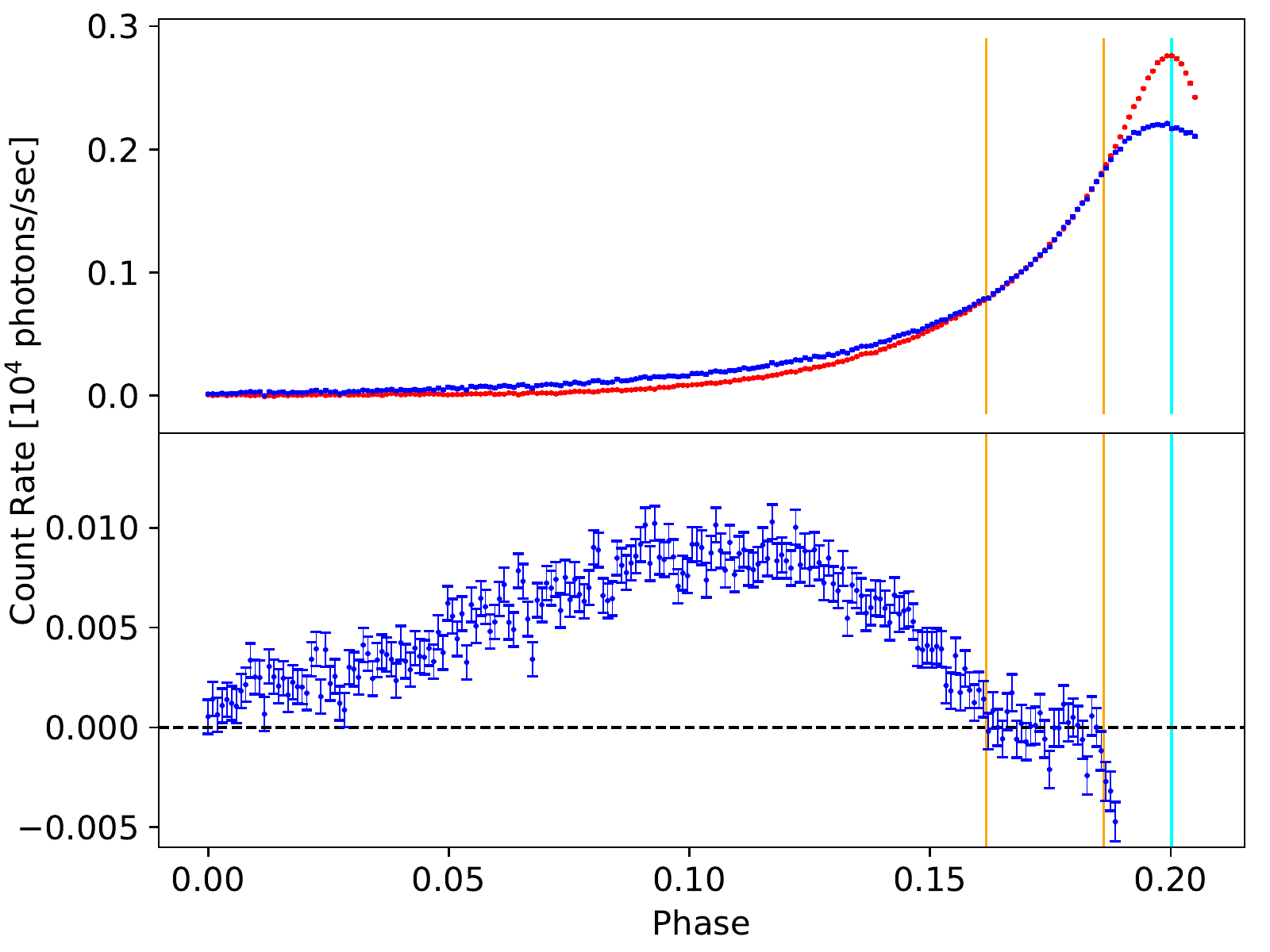}
\vskip-0.2cm
\caption{
	Top panel: Similar to Fig.~\ref{fig2} but after baseline removal, area 
	normalization, shifting the blue curve by $\phi_0$ in phase, and 
	multiplying the red curve by $\kappa$. Bottom panel: Difference of the 
	blue and red curves. The orange lines represent $\phi_1$ and $\phi_2$.
        }
\label{fig3}
\end{figure}

Fig.~\ref{fig3} shows the best result when no smoothing is done. First the complete red 
and blue curves of Fig.~\ref{fig2} (this includes the portions not plotted) are adjusted 
to have zero photons/sec in their first sample; this will be known as baseline removal. 
Next the area under the red curve is made equal to that under the blue curve. Then the 
blue curve is shifted with respect to the red curve by minus one phase sample ($\phi_0 
= - 1/1024$ cycles). Finally the red curve 
is multiplied by $\kappa = 0.977$, and the range of matching is chosen to be $\phi_1 = 
0.1616$ and $\phi_2 = 0.1860$. This data is plotted in the top panel of Fig.~\ref{fig3}. 
The cyan line represents the new peak alignment that includes $\phi_0$ and the vertical 
orange lines represent $\phi_1$ and $\phi_2$. The difference of the blue and red curves 
is plotted in the bottom panel of Fig.~\ref{fig3}. The $25$ data between $\phi_1$ and 
$\phi_2$ have a mean value of $-0.78 \pm 1.8$ photons/sec, the reduced $\chi^2$ being 
$0.93$.  Significantly better results are obtained if two outlying data at phases 
$0.1748$ and $0.1826$ are ignored; then the mean of the $23$ data is $1.12 \pm 1.37$ 
and the reduced $\chi^2$ is $0.51$.

The analysis of this section begins with obtaining an initial set of the parameters
$\phi_0$, $\phi_1$, $\phi_2$ and $\kappa$ by trial and error, for which the reduced
$\chi^2$ is acceptable ($ \le 1.0$), ensuring that the initial range $\phi_2 - \phi_1$ 
is as large as possible. Then three additional phase samples on either side of the 
initial range are utilized to create $16$ phase ranges for searching. So the start of 
the range would have any of the four phases $\phi_1 - n / 1024, n = 0, 1, 2, 3$, while 
the end of the range would have any of the four phases $\phi_2 + n / 1024, n = 0, 1, 2, 
3$. For each of these $16$ ranges $21$ values of $\kappa$ are tried within the range 
$\pm 0.01$ about its initial value in units of $0.001$. Statistics of the difference 
between the red and blue curves of Fig.~\ref{fig3} are obtained for each of these $336$ 
combinations, both with and without the two outlying data.

To understand the reduced $\chi^2$ mentioned above, let $R_i$ and $B_i$ be the count
rates of the red and blue curves respectively in the $i^{th}$ sample in Fig.~\ref{fig2}.
Let $r_i$ and $b_i$ be the corresponding errors, obtained from Poisson statistics and
scaling, and let $R_0$ and $B_0$ be the count rates in the first sample of each curve;
note that the red and blue curves in Fig.~\ref{fig2} have $254$ and $373$ samples
although a smaller amount of this data is plotted in the figure. Let $A_r = \sum_i (R_i -
R_0)$ and $A_b = \sum_i (B_i - B_0)$ be the sum of all red and blue samples respectively
in Fig.~\ref{fig2}, after removing the baseline from each curve. The blue curve in the
top panel of Fig.~\ref{fig3} is $B_{i + 1} - B_0$ where the subscript $i + 1$ represents
shift by minus one sample, while the red curve is $(R_i - R_0) \kappa A_b / A_r$ after
area normalization. The difference between the blue and red curves in the $i^{th}$ sample
is $D_i = (B_{i + 1} - B_0) - (R_i - R_0) \kappa A_b / A_r$.

{{
Its error is $d_i = [b^2_{i + 1} + r^2_i (\kappa A_b / A_r)^2]^{1/2}$. Although $B_0$ and 
$R_0$ are counts in some phase samples, and have Poisson statistical errors on them, they 
do not contribute to $d_i$ because they are essentially constants for our purpose. To 
understand this, one should recall that each $d_i$ can be modeled as $\left [ < D^2_i > 
- < D_i >^2 \right ]^{1/2}$ where $<>$ represents ensemble averaging. The set of ensemble 
$D_i$ for each phase sample $i$ are of course not available, since one has only one $D_i$ 
measurement for each $i$. However this formula brings out the fact that adding or 
subtracting a constant value to a statistical sample changes its mean but not its variance.

However this would not be the case if each phase sample $i$ had a different $B_0$ and $R_0$. 
Then their mean values will bias $D_i$ and not contribute to $d_i$ as before; but their
statistical errors will contribute to $d_i$. For illustration consider observations of radio 
spectral lines using N radio frequency channels, in which observations are first done by 
centering the N channels on the spectral line plus continuum, then observations are done by 
centering them only on the continuum (say by frequency switching). The latter data are
subtracted from the former channel by channel to obtain the spectral line, the noise on
which depends upon the noise on the off-line channels also.

In summary, $B_0$ and $R_0$ behave like constants in the formula for $D_i$, being the same 
for all phase samples $i$. Their statistical errors bias each $D_i$ in the same manner and 
do not contribute to the corresponding $d_i$. Instead, if $B_0$ and $R_0$ were different 
for each phase sample $i$, then their errors would be reflected in each $b_i$. 
}}

Both $D_i$ and $d_i$ are plotted in the bottom panel of Fig.~\ref{fig3}.
The reduced $\chi^2$ is obtained by the formula $\sum_i^N (D_i / d_i)^2 / N$ where the 
sum is over all of the $N$ phase samples between $\phi_1$ and $\phi_2$. The error on the 
reduced $\chi^2$ is $\sqrt{2/N}$.

The red and blue curves of Fig.~\ref{fig2} have a good match in the phase range $\phi_2 
- \phi_1 = 0.1860 - 0.1616 = 0.0244$ after baseline removal, area normalization and phase 
shift; but this is a very small fraction of the phase range over which the reflection
symmetry is implied. The bottom panel of Fig.~\ref{fig3} shows a broad and faint 
emission component of amplitude $\approx 100$ photons/sec and width $\approx 0.1$ in phase, 
centered at phase $\approx 0.1$ beyond the second peak of the FLC in Fig.~\ref{fig1}, or 
equivalently before the second peak in the reflected FLC in Fig.~\ref{fig2}. Only if this
component is real can the reflection symmetry be believed to exist over a substantial
phase range of $\approx 0.2$.

\section{Results of analysis with symmetric smoothing}

\begin{figure}
\centering
\advance\leftskip-0.3cm
\includegraphics[width=9.2cm]{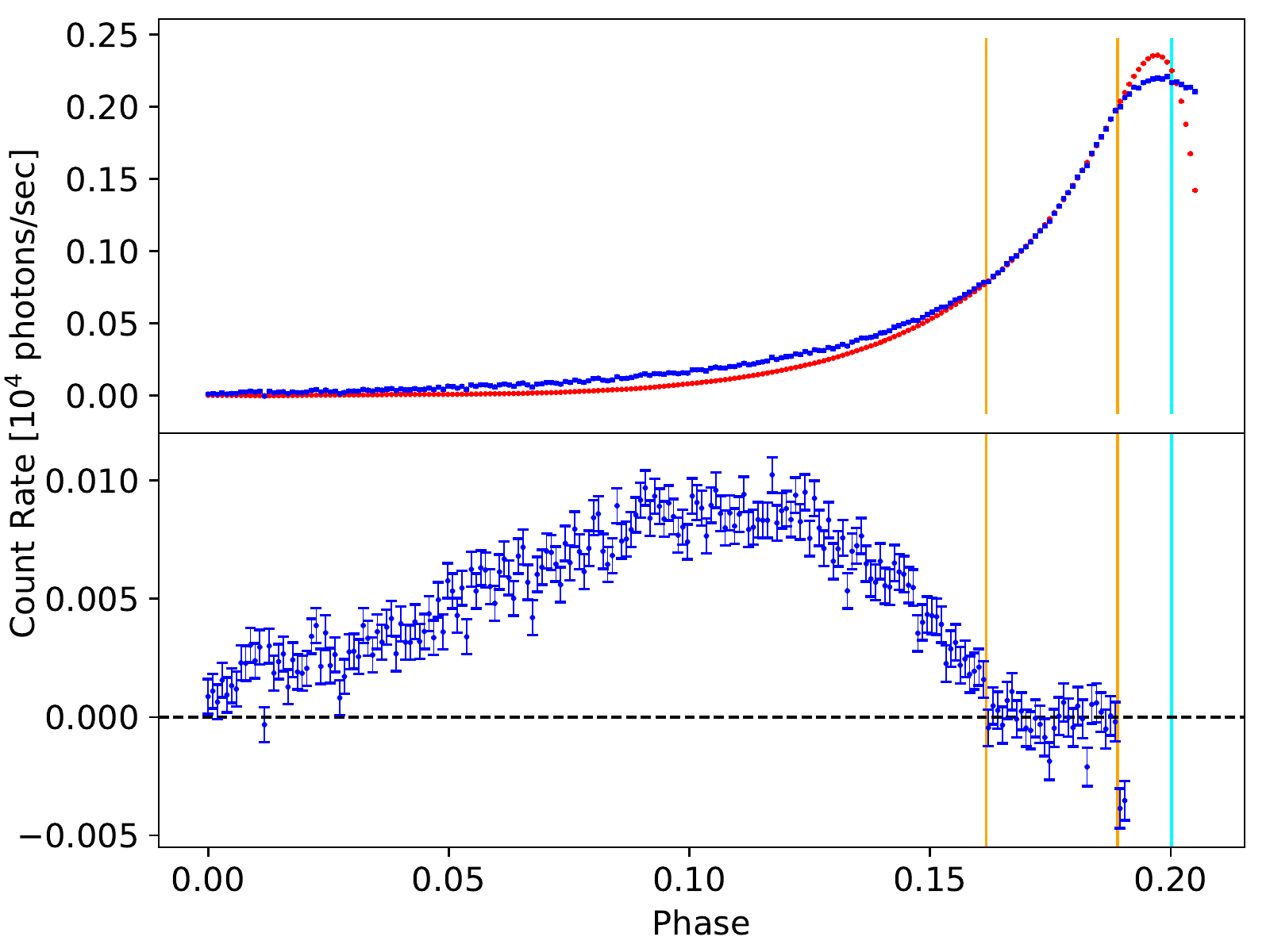}
\vskip-0.2cm
\caption{
	Similar to Fig.~\ref{fig3}, except that the red curve here is the red
	curve of Fig.~\ref{fig3} after being convolved with a Laplace function.
        }
\label{fig4}
\end{figure}

Three symmetric smoothing functions were tried -- Gaussian, Lorentzian and Laplace, 
the last being two exponential functions set back-to-back, its functional form being
$f(x) = \exp(-\vert x \vert / \sigma) / (2 \sigma)$. The Laplace smoothing gave the 
best results that are shown in Fig.~\ref{fig4}; Gaussian and Lorentzian smoothing
fared poorly in comparison.

The initial analysis of this section is similar to that of the previous section --
removal of baseline of the red and blue curves, then their areas being made equal,
then shifting the blue curve by minus one sample in phase ($\phi_0 = - 1/1024$).
Then the red curve is convolved with a Laplace function having $\sigma = 0.0039$. 
Now the convolution theorem states that convolution of two functions implies multiplying 
their Fourier transforms \citep{Bracewell2000}. So the red curve and a sampled 
Laplace function are set into zero padded data arrays $1024$ locations long, and the 
product of their complex Fourier transforms (obtained using FFT) is inverse Fourier 
transformed to obtain the smoothened curve. The smoothened curve is multiplied by 
$\kappa = 0.959$ and the range of matching is chosen to be $\phi_1 = 0.1616$ and 
$\phi_2 = 0.1890$. This data is plotted in the top panel of Fig.~\ref{fig4}.

The search space now consists of five parameters: $\phi_0$, $\phi_1$, $\phi_2$, $\kappa$
and $\sigma$. $\phi_0$ is set to minus one sample ($-1/1024$) as earlier. The $16$
combinations of ($\phi_1, \phi_2$) are coupled with $41$ values of $\kappa$ between
$0.94$ and $0.98$, and $60$ values of $\sigma$ between $0.0012$ and $0.0071$, yielding
$39360$ estimations of the $\chi^2$.

Smoothing the red curve reduces the noise on it. Using the
terminology of the previous section, the count rate in the $i^{th}$ sample of the 
red curve is $C_i = (R_i - R_0) A_b / A_r$; its error is $c_i = r_i A_b /
A_r$. Convolution is essentially a weighted average where the sum of the weights is
unity: $F_i = \sum_{j=-N}^N w_{i-j} C_{i-j} / \sum_{j=-N}^N w_{i-j} = \sum_{j=-N}^N
w_{i-j} C_{i-j}$; the weights $w_{i-j}$ are samples of the Laplace function that is
centered on the $i^{th}$ sample, and $2N$ is some effective number of samples. The
variance on $F_i$ is given by $f^2_i = \sum_{j=-N}^N w^2_{i-j} c^2_{i-j}$ which is
merely the convolution of the scaled variance on the red curve $c^2$ with the square
of the Laplace function $w^2$, after proper normalization for the area of the latter.
So $f_i$ can also be derived using the convolution theorem. The $f_i$ thus estimated
in the Fourier domain has been verified by comparing it with the $f_i$ computed
explicitly in the time domain using the weighted sum formula above. $F$ and $f$ are
plotted as the red curve and its error in Fig.~\ref{fig4} after multiplication by 
$\kappa$.

The phase range of matching has now increased to $0.0274$ in Fig.~\ref{fig4}. The 
mean value of these $28$ data is $-1.23 \pm 1.30$ photons/sec, the reduced $\chi^2$ 
being $0.76$. If the above mentioned two outlier data are ignored the corresponding 
numbers are $0.20 \pm 0.92$ and $0.35$ respectively, which is a significant 
improvement. So Laplace smoothing of the red curve increases the phase range of 
curve matching by three phase samples, as well as improves the statistical similarity 
of the 
two curves. Once again the difference between the red and blue curves is a broad and 
faint component similar to that seen in the bottom panel of Fig.~\ref{fig3}.

Therefore, while smoothing with the Laplace function improves the match between the 
red and blue curves in Fig.~\ref{fig4}, this is still a very small fraction of the 
phase range over which the reflection symmetry is implied.

\section{Results of analysis with asymmetric smoothing}

\begin{figure}
\centering
\advance\leftskip-0.3cm
\includegraphics[width=9.2cm]{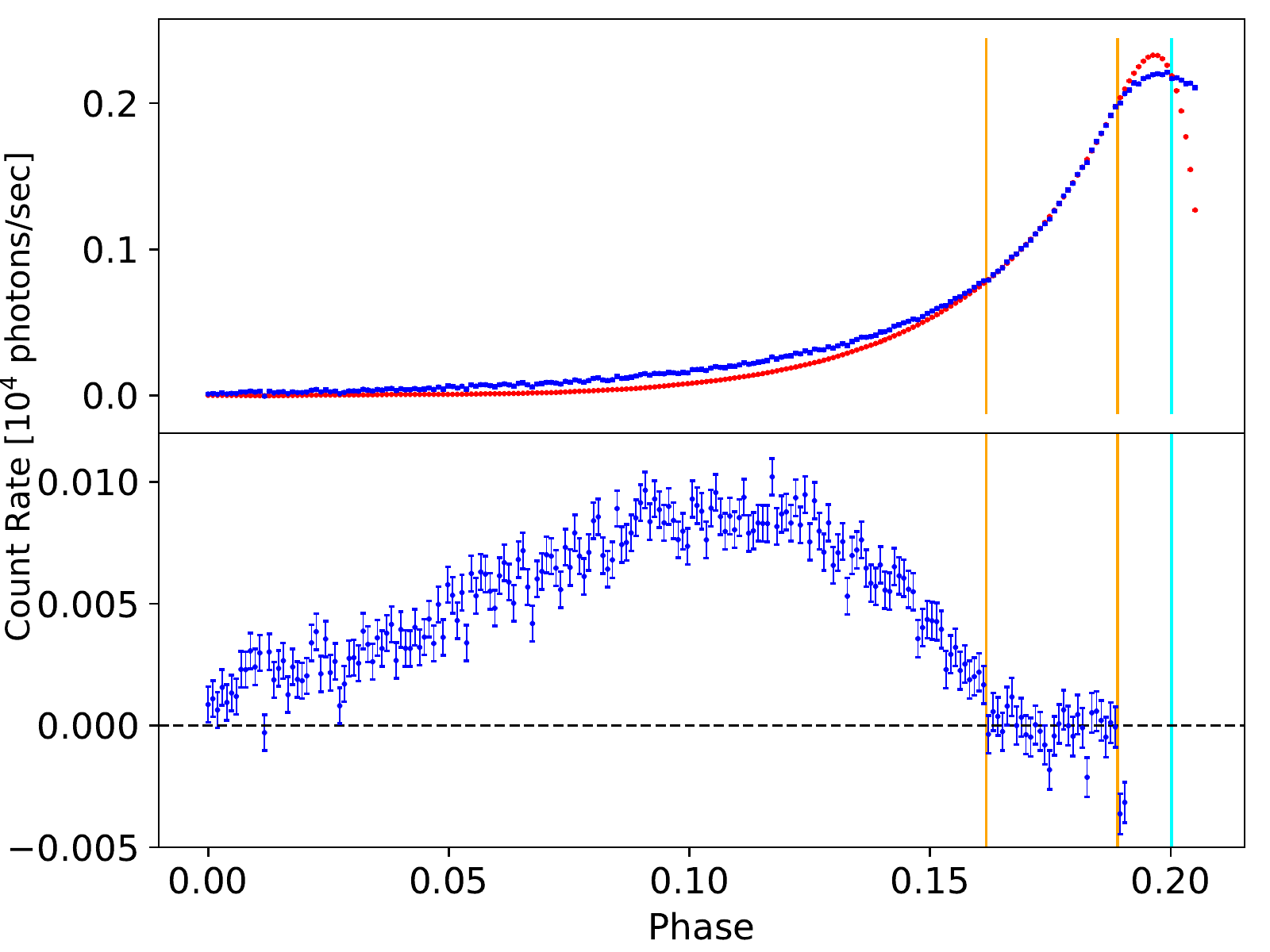}
\vskip-0.2cm
\caption{
	Similar to Fig.~\ref{fig4}, except that the red curve here is the red curve 
	of Fig.~\ref{fig3} after being convolved with an asymmetric Laplace function.
        }
\label{fig5}
\end{figure}

The analysis of the previous section was repeated by choosing asymmetric versions of
the Gaussian, Lorentzian and Laplace functions for smoothing the red curve of 
Fig.~\ref{fig2}. Asymmetric functions were tried to obtain better matching closer to 
the peaks than was obtained in Fig.~\ref{fig4}, but this did not succeed. 
Fig.~\ref{fig5} is an example of one of the best results obtained in this section, 
and is similar to Fig.~\ref{fig4} except that the Laplace function has width $\sigma 
= 0.0039$ at negative phases with respect to its peak and width $\alpha \times \sigma$ 
at positive phases with $\alpha = 0.8$. In Fig.~\ref{fig5} $\phi_0 = - 1/1024$, 
$\phi_1 = 0.1616$ and $\phi_2 = 0.1890$ as in Fig.~\ref{fig4} but $\kappa = 0.936$. 

The
search space now consists of six parameters: $\phi_0$, $\phi_1$, $\phi_2$, $\kappa$,
$\sigma_{-}$ and $\sigma_{+}$. As usual $\phi_0$ was set to minus one sample
($-1/1024$). The $16$ combinations of ($\phi_1, \phi_2$) were coupled with $41$
values of $\kappa$ and $60$ values of $\sigma_{-}$ and six values of $\alpha$ where
$\sigma_{+} = \alpha \times \sigma_{-}$, $\alpha = 0.9 - 0.5$ and $0.1$, yielding
$16 \times 41 \times 60 \times 6 = 236160$ estimations of the $\chi^2$. The ranges
of $\kappa$ had to be chosen differently for different values of $\alpha$.

The phase range of matching $0.0274$ is similar to that in Fig.~\ref{fig4}. The 
mean value of these $28$ data is $-0.76 \pm 1.31$ photons/sec, the reduced $\chi^2$ 
being $0.76$. If the above mentioned two outlier data are ignored the corresponding 
numbers are $0.70 \pm 0.91$, and $0.35$ respectively. The results of Fig.~\ref{fig5}
are similar to those of Fig.~\ref{fig4} since moderate asymmetry is imposed upon
the Laplace function. Asymmetric versions of Gaussian and Lorentzian functions fared 
poorly in comparison. The difference between the red and blue curves in Fig.~\ref{fig5}
is similar to that in Fig.~\ref{fig4} or Fig.~\ref{fig3}.

An asymmetric version of the Gaussian or Lorentzian was derived by multiplying with 
the odd function $1 + \beta \phi$ where $\phi$ is the phase and the magnitude of 
$\beta$ can be chosen to control the level of asymmetry, while its sign controls the 
sense of the asymmetry. Any other odd function can be used such as $1 + \beta \phi^3$. 
The asymmetry increases monotonically with $\beta$. Since the width of the smoothing 
functions is required to be of the order of $\approx 0.004$ for our data, large values 
of $\beta$ are required to achieve the required levels of asymmetry. This makes the 
smoothing function negative at larger $\vert \phi \vert$ values. These negative
function values have to be removed but this makes the smoothing function discontinuous
in its derivative. 

\section{Discussion}

{{
In the introduction it was stated that comparison of the falling edge of the first peak and 
the rising edge of the second peak is model dependent due to the bridge emission. This is 
because the emission between the two peaks in Fig.~\ref{fig1} has to be modeled, at the 
minimum, as a falling edge of the ﬁrst peak (say C1), a bridge emission (C2) and a rising 
edge of the second peak (C3), if one believes that there are two peaks in the FLC and if
one is trying to study their mirror symmetry. Now, Fig.~\ref{fig1} gives the sum C1 + C2 + 
C3 between the two peaks, from which one has to obtain C1 and C3 to study their mirror 
symmetry. Clearly one has to model C1 and C2 to get C3, or C2 + C3 to get C1; either way 
the bridge emission C2 has to be modeled.

On the other hand, the visually suspected mirror symmetry discussed in this work is 
between the observed rising edge of the ﬁrst peak and the observed falling edge of the 
second peak, irrespective of how many independent emission regions of the pulsar's 
magnetosphere contribute to them, not to mention emission from the two independent 
magnetic poles also. So this work does not need to model these two data sets -- it 
uses them directly.
}}

\begin{figure}
\centering
\advance\leftskip-0.3cm
\includegraphics[width=9.2cm]{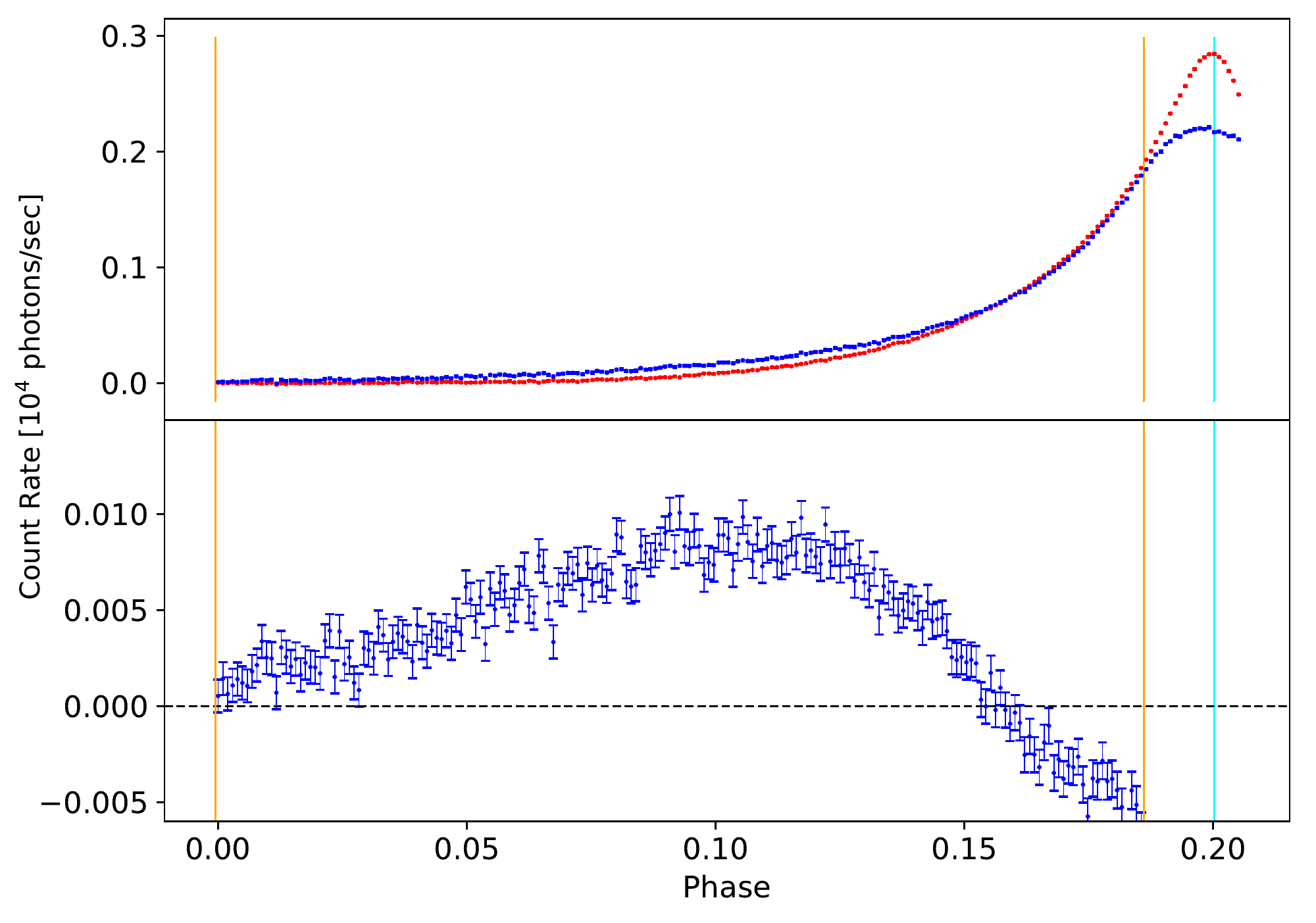}
\vskip-0.2cm
\caption{
	Similar to Fig.~\ref{fig3}, except that the "entire $\chi^2$" is estimated
	in the phase range $\phi_1 = 0.0$ and $\phi_2 = 0.1860$, and $\kappa =
	1.006$.
        }
\label{fig6}
\end{figure}

{{
An important point in this work is that it compares the red and blue curves when they
are aligned at their peaks (almost!). This adds greater credence to any matching
between the two curves, since one is not comparing arbitrary sections of two curves.

A different analysis to explore is when the phase range of comparison in
Fig.~\ref{fig3} becomes $\phi_1 = 0.0$ and $\phi_2 = 0.1860$, which is approximately 
the phase range over which the reflection symmetry has been visually perceived; i.e., 
one wonders what happens when one attempts to minimize the reduced $\chi^2$ over 
almost entire phase range in Fig.~\ref{fig3}. To differentiate from earlier results 
let this be called "entire $\chi^2$".
The "entire $\chi^2$" values are $47.03, 44.43, 42.47, 41.14, 40.43, 40.32, 40.81, 
41.89, 43.54, 45.75$ and $48.51$, for values of $\kappa$ ranging from $0.96$ to
$1.06$ in steps of $0.01$. At the specific value of $\kappa = 0.977$ in Fig.~\ref{fig3}
the "entire $\chi^2$" is $42.99$, while its minimum value is $40.29$ at $\kappa = 
1.006$. There is not much difference between "entire $\chi^2$" values of $42.99$
and $40.29$.

Fig.~\ref{fig6} is similar to Fig.~\ref{fig3} except that the phase range of 
comparison is much larger and $\kappa = 1.006$. The broad and faint component below
$\phi_1 = 0.1616$ persists, while the red and blue curves depart significantly from 
each other in the original phase range of comparison $\phi_1 = 0.1616$ and 
$\phi_2 = 0.1860$. This trend continues for $\kappa$ values departing significantly 
from $1.006$ -- the two curves depart dramatically from each other in the original 
phase range of comparison, while the broad and faint component persists without
significant changes. I believe this is because changes in $\kappa$ make much larger 
changes to the original matching phase interval, in which the light curve is changing
almost exponentially, and make relatively smaller changes in the rest of the phase
range in which the light curve is changing far more slowly.

The "entire $\chi^2$" analysis is probably for those who are interested in finding 
out what sort of minimal emission components exist on the falling edge of the second 
peak, probably for theoretical modeling. The motivation/philosophy of this work is 
quite different. It looks for the largest phase range of matching of the blue and red 
curves, not for the smallest statistical difference between the two.
}}

The main result of this work is that strictly the visually apparent reflection symmetry 
in the FLC of the Crab pulsar is restricted to a very small phase range of $\approx 
0.0274$, which is so small that one can not rule out a coincidental matching. The only
way this can be extended to a substantial phase range is by invoking a broad and faint 
emission component beyond the second peak, that is shown in the bottom panels of figures
\ref{fig3}, \ref{fig4} and \ref{fig5}. Conversely, if in future the spectrum of this 
component turns out to be different from the rest of the FLC, then greater credence can 
be given to its existence; then one can believe that the reflection symmetry in the Crab 
pulsar's FLC extends over a substantial phase range. Pending such confirmations, if at 
all, the rest of this section will speculate on the consequences of supposing that these 
two facts are true, because currently it is too early to reject the existence of the 
reflection symmetry.

\subsection{Mirror symmetry}

{{
Throughout this work the stress has been on the "approximate reflection symmetry" between 
the rising edge of the first peak and the falling edge of the second peak of the Crab 
pulsar's soft X-ray FLC. I believe earlier astronomers (both observers and theorists) also 
implied the same since theirs was essentially a visual observation. The concept of 
perfect reﬂection symmetry is non existent in the current work which is observational. 
Matching of segments of FLCs is subject to the noise on those segments and to the 
phase resolution of the data -- today's perfect matching may be overturned by tomorrow's 
higher quality data that may have less noise and higher phase resolution. The important 
point here is that earlier astronomers were impressed with even the semblance of a 
reasonable reﬂection symmetry since the two edges are supposed to arise from widely 
different regions of the pulsar's magnetosphere. Clearly earlier astronomers were
reporting the zeroth order visual effect while visually smoothing out any higher order
departures from perfect reflection symmetry.
}}

Very early \cite{Cheng1986} and \cite{Romani1995} demonstrated the formation of double peaked
$\gamma$ ray FLCs of RPPs using only a dipole magnetic field, special relativistic aberration 
and light travel time of a photon, without invoking the details of the radiative processes.
It will be very convenient if the mirror symmetry of the two peaks is also explained using
the above minimal inputs. Indeed figure~$5$ of \cite{Romani1995} and Fig.~$8$ of 
\citet{Cheng2000} show simulated light curves that rise steeply towards the first peak and 
fall as steeply after the second peak, but their phase resolution was much worse than that of 
this work. A priori one could suppose that the mirror symmetry can be explained more easily
if the emission was from only one pole of the Crab pulsar instead of two poles, since there
would be one parameter less to deal with. However, \citet{Tang2008} show in their Figure~$6$ 
that the rising edge of the first peak of the Crab pulsar's high energy FLC has significant 
contributions from both of its magnetic poles, while the falling edge of the second peak 
has significant contribution from only one pole. Moreover they invoke emission asymmetry 
between the two poles of the Crab pulsar since the dipole may not be at the center of the 
pulsar. These additional details may make the explanation for the mirror symmetry more
complicated. So this mirror symmetry, if at all it exists, may be an important constraint 
for modeling the soft X-ray FLC of the Crab pulsar.

Several authors have simulated high energy FLCs of RPPs for diverse magnetic field
structures and acceleration gaps under different magnetospheric and geometric assumptions 
\citep{Venter2009, Watters2009, Romani2010, DeCesar2013, Kalapotharakos2014}; these are 
known as light curve atlases. Some of these simulated FLCs do display some sort of 
reflection symmetry between the two peaks, although there are equal number which display only 
translation symmetry. These simulated FLCs depend critically upon the shape and size of 
the emission zone in the magnetosphere and the intensity weight for each point in this 
zone, i.e., the relative amount of radiation emitted from each point. Most authors make 
the simple assumption the intensity weight is uniform across the emission zone. The
mirror symmetry may provide constraints upon the shape and size of the 
emission zone and the intensity weight within it.

That the mirror symmetry improves upon smoothing the data with a Laplace function is
a surprising result; intuitively one might have expected a more common function like 
the Gaussian, which however fares poorly in comparison. This may be yet another critical 
constraint for the modeling of soft X-ray FLCs of RPPs.

\subsection{Broad and faint emission component}

To the best of my knowledge the broad and faint emission component in the Crab pulsar's 
soft X-ray FLC has been reported for the first time in this work. 
{{ It is not surprising that it was undetected so far. In Fig~\ref{fig1} the second 
peak occurs at phase $0.6431$ and the peak of the broad and faint emission component
occurs at phase $\approx 0.6431 + 0.1 \approx 0.7431$, where the FLC value is $\approx
7176$ photons/sec. Clearly a maximum FLC enhancement of $\approx 100 / 7176 \approx 
1.4$\% is not discernible to the human eye in Fig~\ref{fig1}, particularly since this 
component is riding on top of a curving wing of the second peak. Moreover, 
}}
the sum of the count rates of this component in the bottom panel of Fig.~\ref{fig4} in 
the phase range $0 - 0.1616$ is $9078.01 \pm 95.64$ photons/sec, while the corresponding 
number for the blue curve in the top panel of that figure is $1196851.36 \pm 94.53$; so 
{{when this component is integrated}} it is 
only a fraction of $0.0076 \pm 0.00008$ or $\approx 0.76$\%~$\pm 0.01$\% of the 
background that it is riding on. Therefore it is not surprising that this component was 
not evident in the earlier soft X-ray FLCs of the Crab pulsar that had lower sensitivity 
and resolution {{and became evident only after the quantitative analysis of this work}}. 
If this component is indeed real then its nature may be revealed when 
its spectrum is obtained, but that may be difficult due to its faintness. Meanwhile we 
can make the following speculations.

The Crab pulsar's radio FLC has seven components \citep{Eilek2016} some of which are 
prominent only at higher radio frequencies (beyond $\approx 5$ GHz). The positions of two 
of these labeled {{\it HFC1}} and {{\it HFC2}} drift in rotation phase as a function of 
observing radio frequency, and appear at higher phases at higher radio frequencies. See 
Fig~$1$ of \citet{Eilek2016} who state that there is no clear sign of these two components 
in high energy FLCs of the Crab pulsar, except for a possible weak detection above 
$\approx 10$ GeV by \citet{Abdo2010a}, who detect a slight enhancement in the FLC in their 
Fig.~$1$ at phase $\approx 0.74$. Now the second peak in their figure is at phase $\approx 
0.39$. So the faint feature detected by \citet{Abdo2010a} lies at phase $\approx 0.35$ 
beyond the second peak, while the feature detected here lies phase $\approx 0.1$ beyond 
the second peak. Moreover, the width of the feature of \citet{Abdo2010a} is much narrower 
compared to the feature here. Therefore it appears unlikely that the two are connected 
although it should be kept in mind that the two measurements are at widely different 
energies.  Nevertheless it is interesting to speculate if the broad and faint feature 
detected here is in some way connected to the radio {{\it HFC1}} and {{\it HFC2}} 
components of the Crab pulsar, particularly if their high energy counterparts also drift 
in phase.

If the spectrum of this broad and faint component turns out to be completely non thermal, 
then it would have to be explained by the standard light curve modeling done at high
energies \citep{Bai2010, Romani2010, Harding2016}. Figure~$6$ of \citet{Tang2008} 
demonstrates a possibility -- a broad and weak extended emission beyond the second peak
that arises due to widening of the azimuthal extension of the outer gap. However their
component decreases monotonically with phase beyond the second peak while the feature
hear peaks at phase $\approx 0.1$ beyond the second peak.

If the spectrum of this broad and faint component turns out to be thermal, 
then there are interesting possibilities -- it would point to a thermal hot spot
on the surface of the Crab pulsar. It would be more interesting if it turns out
to be like the spectrum of a magnetar, which is a combination of a thermal and two
non thermal components, the former arising from a surface hot spot while the latter
is due to the magnetic atmosphere of the pulsar; see Fig.~$3$ of \citet{Mereghetti2015}, 
Fig.~$5$ of \citet{Kaspi2017} and Fig.~$2$ of \cite{Esposito2020}. This would indicate 
that there is a bit of a magnetar behavior in the Crab pulsar. This would be consistent 
with the assertion of \cite{Esposito2020} in their final remarks that magnetars can
be radio pulsars and ordinary radio pulsars can behave like magnetars. This would also 
be consistent with the relatively high rate of glitch and timing noise activity in the 
Crab pulsar, which is a typical magnetar behavior \citep{Turolla2015, Kaspi2016, 
Esposito2020}.

\section{Data availability}

The data underlying this article are available at the 
NICER\footnote{https://heasarc.gsfc.nasa.gov/docs/nicer/} observatory site at HEASARC
(NASA). The data of Fig.~\ref{fig1} is available in the article's online supplementary 
material.

\section*{Acknowledgments}

I thank the referee for useful comments.

\end{document}